\documentclass[a4paper]{jpconf}
\usepackage{graphicx}
\begin{document}
\title{Heavy Mesons in Nuclear Matter and Nuclei}

\author{Laura Tolos$^{1,2}$, Daniel Cabrera$^2$, Carmen Garcia-Recio$^3$, Raquel Molina$^4$, Juan Nieves$^5$, Eulogio Oset$^5$, Angels Ramos$^6$, Olena Romanets$^7$,  Lorenzo Luis Salcedo$^3$ and Juan M. Torres-Rincon$^8$}

\address{$^1$Instituto de Ciencias del Espacio (IEEC/CSIC), Campus Universitat 
Aut\`onoma de Barcelona, Facultat de Ci\`encies, Torre C5, E-08193 Bellaterra 
(Barcelona), Spain}

\vspace{0.1cm}

\address{$^2$Institut f\"ur Theoretische Physik and Frankfurt Institute for Advanced Studies, Johann Wolfgang Goethe University, Ruth-Moufang-Str. 1,
60438 Frankfurt am Main, Germany} 

\vspace{0.1cm}

\address{$^3$Departamento de F{\'\i}sica At\'omica, Molecular y Nuclear, and Instituto Carlos I de F{\'i}sica Te\'orica y Computacional,
Universidad de Granada, E-18071 Granada, Spain}

\vspace{0.1cm}

\address{$^4$Department of Physics,
Columbian College of Arts $\&$ Sciences,
Corcoran Hall,
725 21st Street NW,
Washington, DC 20052, USA}

\vspace{0.1cm}

\address{$^5$Instituto de F{\'\i}sica Corpuscular (centro mixto CSIC-UV),
Institutos de Investigaci\'on de Paterna, Aptdo. 22085, 46071, Valencia, Spain}

\vspace{0.1cm}

\address{$^6$Departament d'Estructura i Constituents de la Mat\`eria,
Universitat de Barcelona,
Mart\'{\i} i Franqu\'es 1, 08028 Barcelona, Spain}

\vspace{0.1cm}

\address{$^7$KVI, University of Groningen, Zernikelaan 25, 9747AA Groningen, The Netherlands}

\vspace{0.1cm}

\address{$^8$Subatech, UMR 6457, IN2P3/CNRS, Universit\'e de Nantes, \'Ecole de Mines de Nantes, 4 rue Alfred Kastler 44307,
Nantes, France}

\ead{tolos@ice.csic.es}
       
\begin{abstract}
Heavy mesons in nuclear matter and nuclei are analyzed within different frameworks, paying a special attention to unitarized coupled-channel approaches. Possible experimental signatures of the  properties of these mesons in matter are addressed, in particular in connection with the future FAIR facility at GSI.
\end{abstract}

\vspace{-0.5cm}

\section{Introduction}
Over the last decades, matter under extreme conditions of density and temperature has been the subject of study  in order to address fundamental aspects of the strong interaction. This study is intimately linked to several experimental programs, such as SIS/GSI, RHIC/BNL, LHC/CERN project and the forthcoming PANDA and CBM experiments at FAIR. In this context, the properties of hadrons with strange and charm content in hot and dense matter are of particular interest. Whereas the properties of mesons with strangeness in dense matter have been analyzed in connection to the study of exotic atoms as well as the analysis of heavy-ion collisions, the properties of mesons with charm are under scrutiny and will play an important role at FAIR.

In this talk we review different approaches to obtain the in-medium properties of the strange  and open-charm  mesons in nuclear matter and nuclei, paying a special attention to coupled-channels unitarized methods. Several experimental scenarios are analyzed, such as heavy-ion collision data on strange pseudoscalar mesons, the photoproduction of strange vector mesons,  the formation of $D$-mesic nuclei and the $D$-meson propagation from RHIC to FAIR energies.

\section{Strange mesons in matter}

\subsection{Strange pseudoscalar mesons: $\bar K$ in matter}

Early works based on relativistic mean-field calculations \cite{Schaffner:1996kv} obtained very deep potentials of a few hundreds of MeVs at saturation density $\rho_0$ for $\bar K$ in matter, in line with the analysis of data on antikaonic atoms using phenomenological models \cite{Friedman:2007zz}. However, antikaonic-atom data tests matter at the surface of the nucleus and, therefore, do not really provide a suitable constraint on the antikaon-nucleus potential at saturation density.  

Later approaches on unitarized theories in coupled channels based on the chiral approach \cite{Lutz,Ramos:1999ku} or on meson-exchange potentials \cite{Tolos01,Tolos02} obtain a much less attractive potential. In these schemes, the attraction is a consequence of the modified $s$-wave $\Lambda(1405)$ resonance in the medium due to Pauli blocking  \cite{Koch} together with the self-consistent consideration of the $\bar K$ self-energy \cite{Lutz} and the inclusion of self-energies of the mesons and baryons in the intermediate states \cite{Ramos:1999ku}. Attraction of the order of -50 MeV at normal nuclear matter density  is then reached \cite{Ramos:1999ku,Tolos01,Tolos02}. Moreover, higher-partial waves beyond $s$-wave \cite{Tolos:2006ny,Lutz:2007bh,Tolos:2008di,dani} become essential for relativistic heavy-ion experiments at beam energies below 2~GeV per nucleon.  

One of the latest unitarized approaches for calculating the $\bar K$ self-energy in symmetric nuclear matter at finite
temperature is that of  Refs.~\cite{Tolos:2008di,dani}. In this model, the $\bar K$ self-energy and, hence, the spectral function are obtained from the  in-medium $\bar K$-nucleon
interaction  in $s-$ and $p-$waves within a chiral unitary approach.  The evolution of the  $\bar{K}$  spectral function with density and
temperature is shown in the l.h.s. of Fig.~\ref{fig:spec-trans},  defined as
\begin{eqnarray}
S_{\bar K}(q_0,{\vec q},T)= -\frac{1}{\pi}\frac{{\rm Im}\, \Pi_{\bar K}(q_0,\vec{q},T)}{\mid
q_0^2-\vec{q}\,^2-m_{\bar K}^2- \Pi_{\bar K}(q_0,\vec{q},T) \mid^2} \ ,
\label{eq:spec}
\end{eqnarray}
where $\Pi_{\bar K}(q_0,\vec{q},T)$ is the $\bar K$ self-energy. The $\bar K$ spectral function
 shows a broad peak that results from a strong mixing between the
quasi-particle peak and the $\Lambda(1405)N^{-1}$ and
$Y(=\Lambda, \Sigma , \Sigma^*)N^{-1}$ $p$-wave excitations. Temperature and density soften the $p$-wave
contributions to the spectral function at the quasi-particle energy.
 
From heavy-ion collisions there has been a lot of activity aiming at extracting the properties of  $\bar K$ in a dense and hot environment \cite{Hartnack:2011cn}. Initial studies have addressed 
the $\bar K$ production in nucleus-nucleus collisions at SIS energies using a transport model with $\bar K$ that were
dressed with the Juelich meson-exchange model \cite{Cassing:2003vz} or analyzing multiplicity ratios \cite{Tolos:2003qj}. However, the question that still remains is to what extend the properties of $\bar K$ mesons are modified in matter.

\subsection{Vector mesons with strangeness: $\bar K^*$ in matter}

With regard to strange vector mesons in the nuclear medium, very little discussion has been made about their properties.  
Recently, the  $\bar K^*$  self-energy in symmetric nuclear matter has been obtained  within the
hidden gauge formalism of Ref.~\cite{tolos10}. Two sources for the
modification of the $\bar K^*$ $s$-wave self-energy emerged in nuclear matter: one associated to the
decay mode $\bar K \pi$ modified by nuclear
medium effects on the $\pi$  and $\bar K$ mesons  (which accounts for
the $\bar K^* N \to \bar K N, \pi Y, \bar K \pi N, \pi \pi Y \dots$ processes, 
with $Y=\Lambda,\Sigma$); and a second one linked to the interaction
of the $\bar K^*$ with the nucleons in the medium (which accounts for the direct
quasi-elastic process $\bar K^* N \to \bar K^* N$, as well as other absorption
channels involving vector mesons and baryons, $\bar K^* N\to V B $),  coming from a unitarized coupled-channel
calculation. Two resonances are generated
dynamically, $\Lambda(1783)$ and $\Sigma(1830)$, which can be identified with
the experimentally observed states $J^P=1/2^-$ $\Lambda(1800)$ and the
$J^P=1/2^-$ PDG state $\Sigma(1750)$, respectively \cite{Oset:2012ap}.

\begin{figure}[t]
\begin{center}
\includegraphics[width=0.35\textwidth,height=0.3\textwidth]{fig6.eps}
\hfill
\includegraphics[width=0.3\textwidth,height=4.cm]{spectral_ksn.eps}
\hfill
\includegraphics[width=0.32\textwidth,height=4.cm]{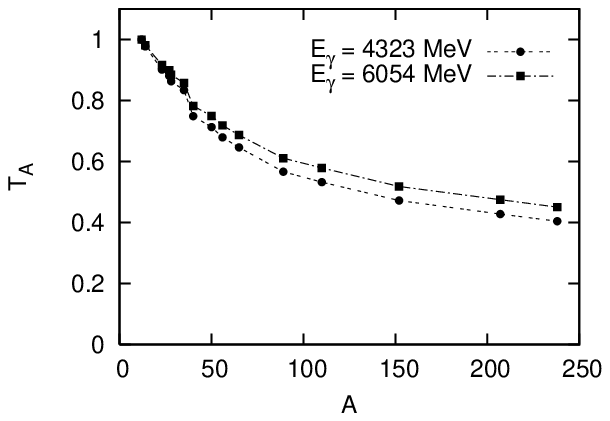}
\caption{Left: $\bar K$ spectral function for different densities, temperatures and momenta \cite{Tolos:2008di}. Middle: $\bar K^*$ spectral function at zero momentum for different densities  \cite{tolos10}. Right: $\bar K^*$ transparency ratio \cite{tolos10}.}
\label{fig:spec-trans}
\end{center}
\end{figure}

 The $\bar K^*$ spectral function at zero temperature is given in a similar way as in Eq.~(\ref{eq:spec}) and is displayed in the middle panel of Fig.~\ref{fig:spec-trans} as a function of the meson energy $q_0$, for zero
momentum and different densities up to 1.5 $\rho_0$. The dashed line refers to
the calculation in free space, where only the $\bar K \pi$ decay channel
contributes, while the other three lines correspond to the self-consistent
calculations,  which incorporate the process $\bar K^* \rightarrow \bar K
\pi$ in the medium, as well as the quasielastic $\bar K^* N \to \bar K^* N$ and
other $\bar K^* N\to V B$ processes.  The structures above the quasiparticle
peak correspond to the dynamically generated $\Lambda(1783) N^{-1}$ and
$\Sigma(1830) N^{-1}$ excitations. Density effects result in a dilution and
merging of those resonant-hole states, together with a  broadening of the
spectral function  due to the increase of collisional and absorption processes. What is clear from the
present approach, is that the spectral function spread of the $\bar K^*$
increases substantially in the medium, becoming at normal nuclear matter density
five times bigger than in free space.

In order to test the $\bar K^*$ self-energy experimentally, we analyze the normalized nuclear transparency ratio, defined as
\begin{equation}
T_{A} = \frac{\tilde{T}_{A}}{\tilde{T}_{^{12}C}}, \hspace{0.5cm} {\rm with} \hspace{0.5cm} \tilde{T}_{A} = \frac{\sigma_{\gamma A \to K^+ ~K^{*-}~ A'}}{A \,\sigma_{\gamma N \to K^+ ~K^{*-}~N}} \ .
\end{equation}
 It describes the loss of flux of $K^{*-}$ mesons in the nucleus and is related to the absorptive part of the $K^{*-}$-nucleus optical potential and, thus, to the $K^{*-}$ width in matter.  We evaluate the ratio between the nuclear cross sections in heavy nuclei and a light one ($^{12}$C), $T_A$, so that other nuclear effects not related to the absorption of the $K^{*-}$ cancel. In the right panel of  Fig.~\ref{fig:spec-trans} we  observe a very strong attenuation of the $\bar{K}^*$
survival probability due to the decay  $\bar{K}^*\to
\bar{K}\pi$ or absorption channels 
$\bar{K}^*N\to \bar K N, \pi Y, \bar K \pi N, \pi \pi Y, \bar K^* N, \rho Y,
\omega Y, \phi Y, \dots$  with increasing nuclear-mass number $A$. This is due
to the larger path that the $\bar{K}^*$ has to follow before it leaves the
nucleus, having then more chances to decay or get absorbed.

\section{Open-charm mesons in matter}

The medium modifications of mesons with charm, such as $D$ and $\bar D$ mesons, have been object of recent theoretical interest  due to the consequences for charmonium suppression.  A phenomenological estimate based on the quark-meson coupling (QMC) model predicts an attractive $D^+$-nucleus potential at $\rho_0$ of $\sim$ -140 MeV \cite{qmc}. The $D$-meson mass shift has also been studied using the QCD sum-rule (QSR) approach \cite{arata,kaempfer}, where a mass shift of -50 MeV at $\rho_0$ for the $D$-meson  has been suggested \cite{arata}.   Recent results on QSR rules for open charm mesons can be found in  \cite{kaempfer}. The mass modification of the $D$-meson is also addressed using a chiral effective model in hot and dense matter \cite{dmeson}, where strong mass shifts were obtained.

With regard to approaches based on coupled-channel dynamics, unitarized methods have been applied in the meson-baryon sector with charm content \cite{Tolos:2004yg,Tolos:2005ft,Lutz:2003jw,Hofmann:2005sw,Mizutani:2006vq,Tolos:2007vh,Molina:2008nh,JimenezTejero:2011fc,Haidenbauer:2007jq,Haidenbauer:2010ch,Wu:2010jy}, partially motivated by the parallelism between the $\Lambda(1405)$ and the $\Lambda_c(2595)$.

More recently, the implementation of heavy-quark spin symmetry (HQSS) has been considered, which is a proper QCD symmetry that appears when the quark masses, such as the charm mass, become larger than the typical confinement scale. The model generates dynamically resonances with negative parity in all the isospin, spin, strange and charm sectors  that one can form from an $s$-wave interaction between pseudoscalar and vector meson multiplets with $1/2^+$ and $3/2^+$ baryons \cite{GarciaRecio:2008dp,Gamermann:2010zz,Romanets:2012hm,Garcia-Recio:2013gaa}.  Within this model, the self-energies and, hence, spectral functions for $D$ and $D^*$ mesons are obtained self-consistently in a simultaneous manner, as it follows from HQSS, by taking, as bare interaction, an appropiately extended WT interaction \cite{Garcia-Recio:2013gaa}. We incorporate Pauli blocking effects and open charm meson self-energies in the intermediate propagators for the in-medium solution  \cite{tolos09}. 

\begin{figure}[t]
\includegraphics[width=0.5\textwidth,height=0.28\textwidth]{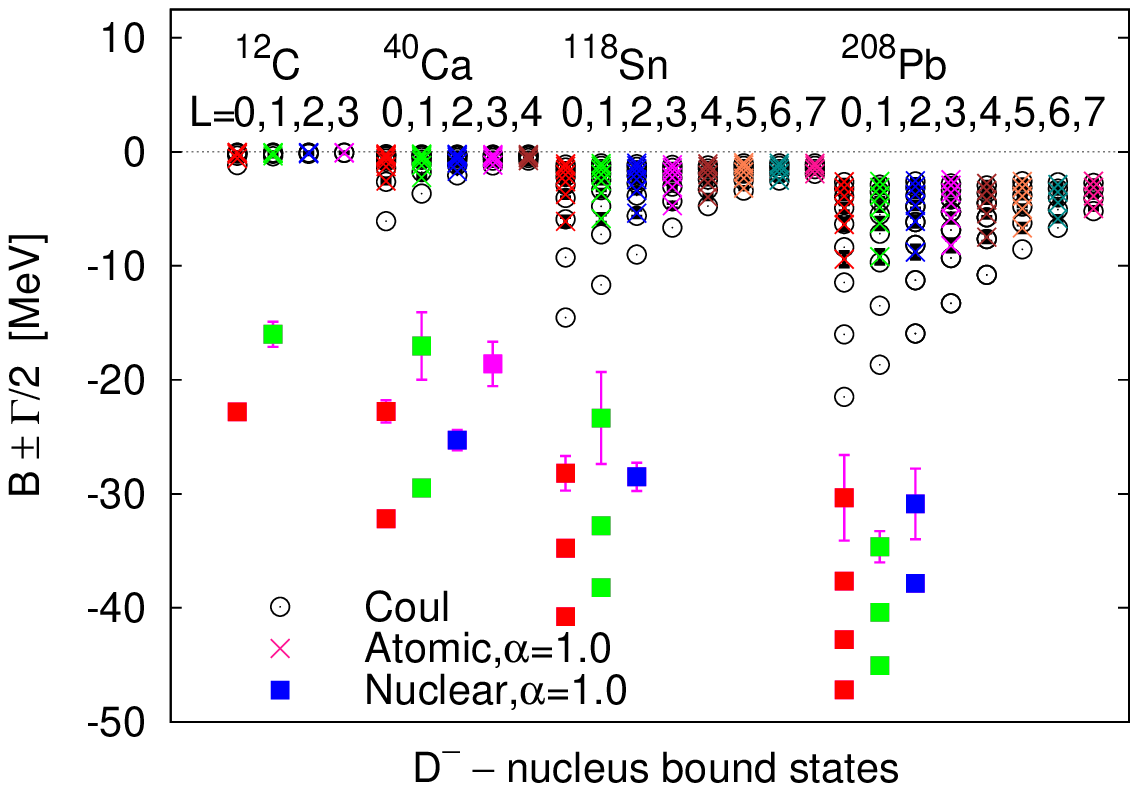}
\includegraphics[width=0.4\textwidth,height=0.25\textwidth]{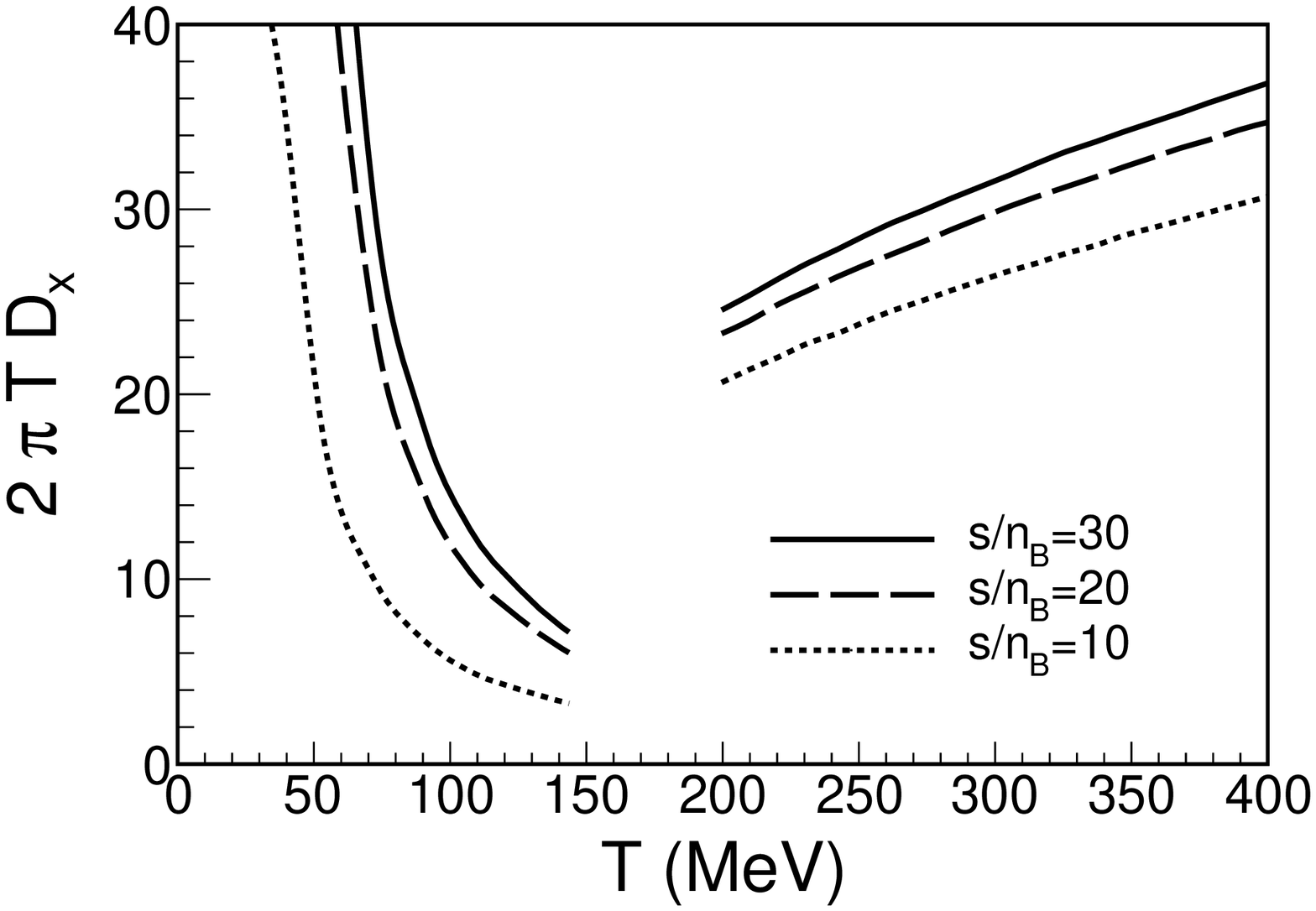}
\caption{Left: $D^-$-nucleus bound states, with $B$ and $\Gamma$  the binding energy and width \cite{GarciaRecio:2011xt}. Right: Spatial diffusion coefficient for $D$ mesons multiplied by $2\pi T$ \cite{Tolos:2013kva}. For recent updates in the high-temperature phase, see Refs.~\cite{Berrehrah:2014tva,Ozvenchuk:2014rpa}. \label{fig3}}
\end{figure}

The detection of the in-medium properties of open charm mesons in matter can be addressed, for example, in the possible formation of $D$ mesic nuclei. The QMC model predicted $D$ and $\bar D$-meson bound states in $^{208}$Pb  relying upon an attractive  $D$ and $\bar D$ -meson potential in the nuclear medium  \cite{qmc}. Within the model that respects HQSS, we obtain that for $\bar D$-mesic nuclei \cite{GarciaRecio:2011xt}, not only $D^-$ (as seen in Fig.~\ref{fig3})) but also $\bar{D}^0$ is bound in nuclei . The spectrum contains states of atomic and of nuclear types for all nuclei for $D^-$  while only nuclear states are present for $\bar{D}^0$ in nuclei. Compared to the pure Coulomb levels, the atomic states are less bound. The nuclear ones are more bound and may present a sizable width. Moreover, nuclear states only exist for low angular momenta.  In what respects to  $D$ mesons \cite{GarciaRecio:2010vt}, $D^0$-nucleus states are weakly bound  in contrast to previous results using the QMC model. Moreover,  those states have significant widths, in particular, for $^{208}$Pb. Only $D^0$-nucleus bound states are possible since the Coulomb interaction prevents the formation of observable bound states for $D^+$ mesons. The experimental detection of $D$ bound states is, however, a difficult task \cite{GarciaRecio:2010vt}.

Another possibility is the study of transport coefficients for a $D$ meson in a hot dense medium composed of light mesons and baryons, such as
it is formed in heavy-ion collisions. One interesting observable is the spatial diffusion coefficient $D_x$ around the phase transition from RHIC to FAIR energies \cite{Tolos:2013kva} (see Fig.~\ref{fig3} right), with a plausible minimum at the phase transition following an isentropic trajectory from the hadronic to the quark-gluon phase
\cite{Berrehrah:2014tva,Ozvenchuk:2014rpa}.

\section*{Acknowledgements}
This research was supported by DGI and FEDER
funds (FIS2011-28853-C02-02,
FIS2011-24149, FIS2011-24154, FPA2010-16963, FPA2013-43425-P) and the
Spanish Consolider-Ingenio 2010 Programme CPAN
(CSD2007-00042), by Junta de Andaluc\' ia Grant
No. FQM-225, by Generalitat Valenciana under Contract
No. PROMETEO/2009/0090, by the Generalitat de Catalunya under
contract 2009SGR-1289, by the EU
HadronPhysics3 project Grant Agreement No. 283286,  Programme "Together" of the Region Pays de la Loire, the FP7-
PEOPLE-2011-CIG under Contract No. PCIG09-GA-2011-291679, BMBF (Germany) under
project No. 05P12RFFCQ, and Ramon y Cajal Programme.

\end{document}